\documentclass[prd,aps,preprint,nofootinbib,epsfig,floats]{revtex4}

\usepackage{graphicx}
\usepackage{times}
\usepackage{epsfig}
\usepackage{amsmath}
\usepackage{amsfonts}
\usepackage{amssymb}
\usepackage{url}
\usepackage{hyperref}
\usepackage{subfigure}
\newcommand{\be}{\begin{equation}}
\newcommand{\ee}{\end{equation}}
\newcommand{\bea}{\begin{eqnarray}}
\newcommand{\eea}{\end{eqnarray}}

\begin{document}

\pagestyle{plain}

\title{$\nu$-GMSB with Type III Seesaw and Phenomenology }

\author{R. N. Mohapatra\footnote{
  e-mail: rmohapat@physics.umd.edu},
       }
\affiliation{Maryland Center for Fundamental Physics and Department of
Physics, University of Maryland, College Park, MD 20742, USA}

\author{Nobuchika Okada\footnote{
  e-mail: okadan@post.kek.jp},}
\affiliation{Maryland Center for Fundamental Physics and Department of
Physics, University of Maryland, College Park, MD 20742, USA }
 \affiliation{Theory Group, KEK,
 1-1 Oho, Tsukuba, 305-0801, Japan}

\author{Hai-Bo Yu\footnote{
  e-mail: haiboy@uci.edu}}
\affiliation{Department of Physics and Astronomy,
 University of California, Irvine, CA 92697, USA}

\date{July, 2008}

\preprint{\vbox{\hbox{UMD-PP-08-012}}}
\preprint{\vbox{\hbox{UCI-TR-2008-30}}}

\begin{abstract}
We show that when the supersymmetric SU(5) model is extended to
explain small neutrino masses by the type III seesaw mechanism, the
new {\bf 24}-dimensional fields needed for the purpose can act as
messengers for transmitting SUSY breaking from a hidden sector to
the visible sector. For the three {\bf 24} case, the constraints of
grand unification and suppressed lepton flavor violation restrict
the seesaw scale in this case to be in the narrow range of
$10^{12}-10^{13}$ GeV. The model predicts (i) a stable LSP gravitino
with mass in the range of 1-10 MeV which can be a cold dark matter
of the universe; (ii) a stau NLSP which is detectable at LHC; (iii)
a lower bound on the branching ratio $BR(\mu \to e \gamma)$ larger
than $10^{-14}$ testable by the ongoing MEG experiment as well as
characteristic particle spectrum different from other SUSY breaking
scenarios. We also discuss the case with two {\bf 24} fields, which
is the minimal case that can explain neutrino oscillation data.

\end{abstract}

\maketitle

\section{Introduction \label{sec1}}
Supersymmetry (SUSY) is considered to be a main ingredient for TeV
scale physics since it resolves several conceptual issues of the
Standard Model (SM) such as (i) gauge hierarchy problem and (ii) electroweak
symmetry breaking while providing a natural candidate for the dark matter
of the universe and predicting unification of forces at a very high
scale.

Supersymmetry must however be broken to be in accord with
observations and understanding the origin and nature of SUSY
breaking is a primary focus of research in particle physics today.
One interesting proposal is the so-called gauge mediated
supersymmetry breaking (GMSB) where SUSY is broken in a hidden
sector and is transmitted via gauge forces to the Standard Model
sector \cite{GMSB}. The transmission of SUSY breaking to the visible
sector is carried out by SM gauge non-singlet vector like pairs of
superfields (known as messenger fields) via their gauge couplings.

There are several aspects of this interesting proposal that one
might like to improve to make it more appealing : (i) the messenger
fields in GMSB models are generally introduced solely for the
purpose of transmitting the SUSY breaking information and play no
other role, which therefore allows considerable freedom in building
models, making it less
easy to test them; (ii) in simple GMSB scenarios, it is generally hard to
understand the magnitude of $B\mu$ term \cite{bmu}; (iii) typical
GMSB models are specified by two arbitrary hidden sector parameters: the
SUSY breaking strength $F$ and messenger mass $M$, with the soft breaking
parameters in the visible sector given by
$\frac{\alpha}{4\pi}\frac{F}{M}$; it would make the model more predictive
if either of
these parameters could be further constrained (or determined) from
independent physics considerations; (iv) in simple GMSB scenarios
where the messenger fields are chosen to be ${\bf 5}\oplus {\bf\bar
5}$ fields under SU(5) group, the messenger and SM matter fields can
mix leading to large flavor changing effects \cite{Dine}. Any new
suggestion  that remedies one or more of these problems is certainly
worth serious investigation.

Most of the above considerations for GMSB are done in the context of
Minimal Supersymmetric Standard Model (MSSM). However, since the
observation of neutrino masses clearly imply extension of MSSM to
include new physics, an important question is to study whether such
extensions can throw any light on the above problems of GMSB. There
exist several such investigations in the literature which show that
neutrino mass physics can impact SUSY breaking
\cite{rossi,sogee,haibo}. We will call the subset of these models
that use neutrino mass physics to implement gauge mediated SUSY breaking
as $\nu$-GMSB models. One
example of this is the work of Ref.~\cite{rossi}, which uses type II
seesaw mechanism \cite{type2}.

 In this paper we discuss an extension of MSSM embedded into SU(5)
GUT where small neutrino masses are explained via the type III
seesaw mechanism \cite{type3}. These models have interesting
phenomenology and have been studied in several recent papers
\cite{type31}.
The basic idea of type III seesaw is to add SM triplet Higgs fields
with zero hypercharge so that they replace the right handed
neutrinos of the type I seesaw \cite{seesaw}. When embedded into
SU(5) GUT theories, these triplets become part of
 new {\bf 24}-dim. Higgs fields that need to be added to the minimal
SU(5). We will consider supersymmetric version of this model since gauge
coupling unification in these models is automatic in the presence of full
SU(5) multiplets.

 The new {\bf 24}-dimensional fields added to minimal SUSY
SU(5) \cite{dimo} play a dual role in the model considered in the
present paper: in addition to making neutrino masses small via type
III seesaw, they also play the role of messenger fields and transmit
the hidden sector SUSY breaking via GMSB. We call this model
$\nu$-GMSB with type III seesaw and study its phenomenological
implications. Typically current neutrino mass observations require
only two {\bf 24} fields; however, these {\bf 24} fields are like
matter fields and having three generations of them has a certain
appeal since each {\bf 24} goes with one generation. So in the bulk of
this paper, we will discuss the model with three {\bf 24}s and its
implications. In a subsequent section, we also discuss the two {\bf 24}
case, where some of the constraints present in the three {\bf 24} case
are more relaxed. For both cases,  we find that
the large dimensionality of messenger fields makes these
models distinguishable
from other types of GMSB models and testable in near
future.

Our main results for the three {\bf 24} fields case, are: (i) the
prediction of the messenger mass in a narrow range of
$10^{12}-10^{13}$ GeV, which precisely is the range to lead to small
neutrino masses; (ii) the prediction of a stable lightest
super-partner (LSP) gravitino with mass in the range of 1-10 MeV
which can be a cold dark matter of the universe if the reheating
temperature after inflation is around $10^5-10^6$ GeV; (iii) a lower
bound of about 200 GeV for slepton masses with stau being the next
to lightest super-partner (NLSP) which is detectable at the Large
Hadron Collider (LHC); (iv) a lower bound on the branching ratio
$BR(\mu \to e \gamma)$ larger than $10^{-14}$ testable by the MEG
experiment as well as characteristic particle spectrum different
from other SUSY breaking scenarios. For the two {\bf 24} case, the
range of allowed messenger mass is between $10^9$ GeV to $10^{13}$
GeV. This case allows for the possibility of gravitino mass anywhere
from few keV to 10 MeV. In the lower mass range, it could be a warm
dark matter. This would require reheating temperature closer to a
TeV.

The paper is organized as follows. In Sec. II, we present
the basic structure of the model and in Sec. III we examine
the upper and lower limits on the messenger mass scale from
gauge coupling unification and lepton flavor violation;
in Sec. IV, we discuss phenomenology and cosmology of this model
such as its predictions for particle spectrum, gravitino dark
matter and in particular we emphasize an important characteristic
prediction of our model that stau is the next LSP (NLSP) which
may be observable at the LHC via its gravitino decay mode;
in Sec. V, we also discuss the case with two {\bf 24} fields,
which the minimal case that can explain neutrino oscillation data;
Sec. VI is devoted to conclusions.

\section{SU(5) model with Type III Seesaw}
Our proposed $\nu$-GMSB is  an extension of supersymmetric
SU(5) model \cite{dimo} to accommodate small neutrino masses via the
type III seesaw mechanism.
It has the following matter $\bar{F}_i ({\bf \bar{5}})$, $T_i({\bf 10})$
(i=1,2,3 for generations) and Higgs fields:  $\bar{F}_H ({\bf
\bar{5}})\oplus  {F}_H({\bf 5})$ and $\Phi({\bf 24})$ as in the minimal
model extended by the addition of three extra {\bf 24}-dimensional
fields, denoted by $\Sigma_i$, (i=1,2,3). The primary role of these
extra fields is to generate small neutrino masses via type III seesaw
mechanism. To illustrate this we write the matter part of the
superpotential as follows:
\begin{eqnarray}
W_m &=& Y_d^{ij}\bar{F}_i \bar{F}_H T_j
  + Y_u^{ij} T_i T_j F_H   \nonumber \\
 &+& Y_{\nu}^{ij} \bar{F}_i {F}_H \Sigma_j
 + M_\Sigma {\rm Tr}{\Sigma^2_i}
\label{Wm}
\end{eqnarray}
In the next section, we will promote $M_\Sigma$ to be the VEV of the
singlet hidden sector field that also breaks supersymmetry.

SU(5) breaking and doublet-triplet splitting is achieved by the
Higgs part of the superpotential given by:
\begin{eqnarray}
W_H=M_\Phi {\rm Tr}(\Phi^2)+\eta {\rm Tr}(\Phi^3)
\end{eqnarray}
Note that the neutrino masses in this case arise via the diagram
in Fig.~\ref{fig1} and are given by the formula similar
to type I seesaw formula \cite{seesaw}:
\begin{eqnarray}
{\cal M}_\nu = -v^2 \sin^2 \beta Y^T_\nu M^{-1}_\Sigma Y_\nu
\end{eqnarray}
with the VEV of the up-type Higgs doublet,
$\langle H_u \rangle = v \sin \beta $, in MSSM.
For $Y_\nu \sim 0.1-1$, it gives neutrino masses in the eV range if
$M_\Sigma \sim 10^{13}$ or so.
We will see that independent considerations such as those from
gauge coupling unification and suppressed flavor violation
indeed constrain this mass to be in the same range of
$10^{12}-10^{13}$ GeV giving some level of uniqueness to the model.

We will prevent couplings between $\Sigma$ and $\Phi$ fields by a $Z_2$
symmetry under which all matter fields (including the $\Sigma$ field) are
odd and all other fields are even. This also restores R-parity as a good
symmetry of the model making the LSP of the model stable making it a
possible dark matter of the Universe as we see below.

\section{Gauge mediation by {\bf 24} ($\Sigma$) fields}
The goal of this paper is to use the same $\Sigma$-fields used to give
small neutrino mass as messengers of SUSY breaking from the hidden
sector. For this purpose, first we set $M_\Sigma =0$ in
Eq.~(\ref{Wm}) and we write the following superpotential:
\begin{eqnarray}
W_{SSB}= \lambda S {\rm Tr} \Sigma^2_i
\end{eqnarray}
We require that $ \langle S \rangle \neq 0$ and $\langle F_S \rangle \neq 0$.
The masses of the fermionic components of $\Sigma$ fields are
given by $ M = \lambda \langle S \rangle $ whereas those of
the scalar components are given by
$M^2_{\Sigma^{\pm}}= M^2 \pm F_S$ where
$\Sigma^\pm=\frac{1}{\sqrt{2}}(\Sigma \pm \Sigma^\dagger)$.

One can now write down the soft SUSY breaking terms. There are two
contributions: gauge contribution and Yukawa contribution \cite{chacko}.
For Yukawa couplings involving the $\Sigma$ fields small compared to the
gauge couplings (e.g. $Y_\nu \sim 0.1$ and $g_{1,2,3} \gtrsim 0.3$),
the gauge coupling contributions dominate and
we find sfermion masses to be given by:
\begin{eqnarray}
 m_{\tilde{f}}^2 (\mu) &=&
 \sum_i  2 c_i \left( \frac{\alpha_i(\mu)}{4 \pi} \right)^2
 \left( \frac{F_S}{M} \right)^2 \; N_m \; G_i(\mu, M) \; ,
\label{scalarmass}
\end{eqnarray}
where
\begin{eqnarray}
G_i(\mu, S) =
\left( \xi_i^2 + \frac{N_m}{b_i} (1-\xi_i^2)
 \right)
\eea
with
\begin{eqnarray}
\xi_i  \equiv  \frac{\alpha_i(M)}{\alpha_i(\mu)}
= \left[ 1+ \frac{b_i}{2 \pi} \alpha_i(\mu) \mbox{ln}
\left( \frac{M}{\mu} \right)  \right]^{-1}  .
\label{xi}
\end{eqnarray}
Here $b_i$ are the beta function coefficients for different groups,
$c_i$ are the quadratic Casimirs, $N_m=15$ is the Dynkin index for
three {\bf 24}-dimensional messenger fields,
and the sum is taken corresponding to the representation
of the sparticles under the SM gauge groups.
For gaugino masses we have
\begin{eqnarray}
  \frac{M_1(\mu)}{\alpha_1(\mu)}
 =\frac{M_2(\mu)}{\alpha_2(\mu)}
 =\frac{M_3(\mu)}{\alpha_3(\mu)}
 = \frac{N_m}{4 \pi}\frac{F_S}{M}.
\label{gauginomass}
\end{eqnarray}

Note that at the messenger scale, there is a hierarchy between the
gaugino mass and sfermion mass because of the large Dynkin index,
$N_m=15$. For example, the ratio of the right-handed slepton mass
and Bino is found to be $m_{\tilde{e^c}}^2(M)/M_1^2(M) =
1.2/N_m=0.08$ at the messenger scale $M$. Therefore, stau is most
likely to be the NLSP in our model.

The gravitino mass is given by $m_{3/2} \sim F_S/M_{Pl}$. In order
to determine these masses, we need to know the values of ratio
$\Lambda\equiv F_S/M$. In simple GMSB models, the value of $\Lambda$
is fixed by the requirement that squark and slepton masses must be
below a TeV so that one does not require fine-tuning to understand
the weak scale. This however does not fix the $F_S$ and $M$
(although avoiding having tachyonic scalar messenger fields gives
$F_S \leq M^2$) individually leaving the gravitino mass pretty much
a free parameter only to be constrained by phenomenology and
cosmology. In the case of our $\nu$-GMSB, however, as we see below,
there are very strict bounds on $M$ from gauge coupling unification
and suppressed flavor violation; therefore gravitino mass is allowed
only within a very narrow range.

\subsection{Gauge coupling unification constraint
 on the messenger scale}
It is well known that gauge coupling unification property of MSSM remains
true in the presence intermediate scale multiplets as long as they
are full multiplets of SU(5) group. In our case we have full {\bf
24}-dim. multiplets above the weak scale so that we maintain the unification.
However, while these extra multiplets leave the GUT scale unchanged,
they increase the magnitude of the value of gauge coupling
at the unification scale, $M_U$. Since it is necessary to maintain
the validity of perturbation theory till above the GUT scale and
preferably to the Planck scale,
this will impose a lower bound on the mass of the new {\bf 24}-fields.

Considering the evolution of the $\alpha_1$
in the presence of the new fields, we find
\begin{eqnarray}
 \alpha^{-1}_1 (M_U)-\alpha^{-1}_1(M_S) =
 -\frac{33}{10 \pi} \ln \frac{M}{M_S}-
  \frac{108}{10\pi} \ln\frac{M_U}{M},
\end{eqnarray}
where $M_S \sim 1$ TeV is a typical soft mass scale, and $M_U \simeq
2 \times 10^{16}$ GeV is the GUT scale. From this equation, we see
that if we want to keep $\alpha^{-1} (M_U) \gtrsim  1$, we get $M
\equiv \lambda \langle S \rangle \gtrsim 6.2 \times 10^{11}$ GeV.
Similar constraints also arise from the $\alpha_{2,3}$ evolution.
Thus this way of implementing gauge mediation implies that we must
have $\sqrt{F_S} \gtrsim 10^{7.7}-10^{8.2}$ GeV for
a typical soft mass 100 GeV -1 TeV. This in turn
implies that the gravitino mass in these models is:
\begin{eqnarray}
 M_{3/2} \gtrsim 1-10 \; {\rm MeV}.
\end{eqnarray}
We show in the next section under what conditions this gravitino can
be a dark matter of the Universe.

\subsection{Upper limit on messenger mass
from lepton flavor violation (LFV)}
In Eq.~(\ref{Wm}), the messenger fields have the Dirac
Yukawa coupling ($Y_\nu$) to $\bar{{\bf 5}}$ matters,
through which flavor-dependent sfermion masses are induced.
For example, off-diagonal elements of left-handed slepton
mass squared at the messenger scale is estimated as
\begin{eqnarray}
\Delta m_{\tilde \ell ij}^2 \sim
 m_{\tilde \ell}^2 \times
 \frac{(Y_\nu^\dagger Y_\nu)_{ij}}{g_2^2},
\label{mij}
\end{eqnarray}
where $m_{\tilde \ell}^2 $ is the flavor-diagonal
soft mass squared from gauge interactions.
When the Yukawa coupling is small, we can neglect
 their RGE evolutions.

In this subsection, we show that type III seesaw combined with
present experimental constraints on lepton flavor violation e.g.
branching ratio for $\mu\to e \gamma$ gives an upper limit on the
messenger mass $M= \lambda \langle S \rangle$. This can be seen
qualitatively looking at the seesaw formula for neutrino masses.
Note that if the messenger scale is higher, one must increase the
Dirac Yukawa coupling $Y_\nu$ in order to get the eV range neutrino
mass to fit neutrino oscillation data. Maximal neutrino mixing then
suggests that off-diagonal elements of $Y_\nu$ must also be large.
This would therefore increase the off-diagonal elements of
left-handed slepton mass and so the LFV branching ratios. The
present experimental upper limit can therefore be used to set an
upper limit on the $Y_\nu$ elements and hence the messenger scale
$M$ so that eV neutrino masses emerge.

To make this argument quantitative, we assume that the three $\Sigma$
masses are degenerate and $Y_\nu$ is a real matrix.
Inverting the seesaw formula, we have
\bea
 Y_\nu^T Y_\nu = - \frac{M}{v^2 \sin^2 \beta}  {\cal M}_\nu
 = - \frac{M}{v^2 \sin^2 \beta} U_{TB} {\cal D}_\nu U_{TB}^T ,
\eea
where ${\cal D}_\nu$ is a diagonal mass eigenvalue matrix,
and we have assumed that ${\cal M}_\nu$ is diagonalized
by the tri-bimaximal mixing matrix,
\bea
U_{TB}= \left(
\begin{array}{ccc}
  \sqrt{\frac{2}{3}} & \sqrt{\frac{1}{3}} &  0  \\
 -\sqrt{\frac{1}{6}} & \sqrt{\frac{1}{3}} &  \sqrt{\frac{1}{2}} \\
 -\sqrt{\frac{1}{6}} & \sqrt{\frac{1}{3}} & -\sqrt{\frac{1}{2}}
\end{array}
\right) .
\eea
In our analysis, we consider two typical mass spectra,
the normal-hierarchical case (${\cal D}_\nu^{NH}$) and
the inverted-hierarchical case (${\cal D}_\nu^{IH}$) such as
\bea
 {\cal D}_\nu^{NH}
  = {\rm diag} (0, \sqrt{\Delta m_{12}^2}, \sqrt{\Delta m_{13}^2}) ,
 \; \;
 {\cal D}_\nu^{IH}
  = {\rm diag}(\sqrt{\Delta m_{13}^2}, \sqrt{\Delta m_{12}^2+\Delta m_{13}^2},
  0)
\label{numass}
\eea
for the neutrino oscillation data \cite{nudata}
\bea
\Delta m_{12}^2 &=& 7.6 \times 10^{-5} \; {\rm eV}^2, \nonumber  \\
\Delta m_{13}^2 &=& 2.4 \times 10^{-3} \; {\rm eV}^2 .
\eea
For example, one $Y_\nu$ texture for the normal-hierarchical case
 for $\tan \beta =10$ is found to be
\begin{eqnarray}
 Y_\nu^\dagger Y_\nu = - \left(
\begin{array}{ccc}
 0.0000969 & 0.0000969  & 0.0000969 \\
 0.0000969 & 0.000914   & -0.000720  \\
 0.0000969 & -0.000720  & 0.000914
\end{array}  \right) \times \left(
\frac{M}{10^{12}\;{\rm GeV}}  \right).
\end{eqnarray}

In our analysis, we adopt an approximate formula of
 the LFV decay rate \cite{Hisano-etal} \cite{earlyLFV},
\begin{eqnarray}
 \Gamma (\ell_i \rightarrow \ell_j \gamma)
  \sim  \frac{e^2}{16 \pi} m_{\ell_i}^5
  \times  \frac{\alpha_2}{16 \pi^2}
  \frac{| \Delta  m^2_{\tilde{\ell} ij}|^2}{m_{\tilde{\ell}}^8}
  \tan^2 \beta .
 \label{LFVrough}
\end{eqnarray}
Using given $Y_\nu$ textures for both the normal- and
inverted-hierarchical cases and Eqs.~(\ref{scalarmass}) and (\ref{mij}),
in Fig.~\ref{fig2} and \ref{fig3}, we plot the branching ratio
$BR (\mu \to e \gamma)$ as a function of the $\Sigma$ mass $M$,
together with the current experimental bound \cite{MEGA1},
$BR (\mu \to e \gamma) \leq 1.2 \times 10^{-11}$.
In the same way, we obtain the branching ratio of LFV tau decay.
Once the $Y_\nu$ texture is fixed, the following ratio
 is determined independently of $M$ and $\tan \beta$:
\bea
\frac{BR (\tau \to \mu \gamma)}{BR (\mu \to e \gamma)}
\simeq 10.3 \; {\rm and} \; 1.74 \times 10^3,
\label{ratio}
\eea
respectively, for the normal- and inverted-hierarchical cases.

We can draw two conclusions by combining the messenger mass
constraints from gauge coupling unification and lepton flavor
violation discussed above: (i) the seesaw scale is in a very
restricted range of $10^{12}-10^{13}$ GeV and (ii) the current MEG
experiment \cite{meg} searching for the process $\mu\to e\gamma$ can
test this model since for the lowest allowed value for messenger
mass $M$, the prediction for the  $BR(\mu\to e \gamma)$ is above
$10^{-14}$ accessible to this experiment. Note that naively one
might think that since the $BR(\mu\to e\gamma)$ scales like
$m_{\tilde{\ell}}^{-4}$, one might reduce this by choosing a higher
value of the slepton mass. However, in the GMSB model like ours,
such increase can come only from an increase of $F_S/M$, which is
fixed by considerations such as Higgs mass fine-tuning. We therefore
do not have much room to reduce the $\mu\to e+\gamma$ branching
ratio. We also note that the $BR(\mu\to e\gamma)$ is predicted to be
lower for the case of inverted mass hierarchy for neutrinos compared
to the normal hierarchy case.

\section{Phenomenology of $\nu$-GMSB}

\subsection{sparticle and Higgs boson mass spectra}

As mentioned before, one point we clearly notice from
the GMSB sparticle mass formulas of Eq.~(\ref{scalarmass})
and (\ref{gauginomass}) is that sfermion masses are
lighter than their corresponding gaugino masses.
This is because of the Dynkin index $N_m=15$ and
sfermion masses are suppressed by a factor $1/\sqrt{N_m}$
compared to corresponding gaugino masses.
This is a similar structure to the no-scale supergravity
\cite{no-scale}.
For example, we find that gluino is the heaviest sparticle.
One of the most characteristic feature in $\nu$-GMSB
with type III seesaw is that the (mostly right-handed) stau
is the NLSP.

In Table~\ref{table1} and \ref{table2}, examples of sparticle and Higgs mass spectra
are presented for $\tan \beta=$10 and 45, respectively%
\footnote{
We have used SOFTSUSY 2.0.11 \cite{softSUSY}
to generate sparticle and Higgs mass spectra.
}.
In these Tables, for comparison, we also show the mass spectra
in GMSB with type II seesaw (corresponding to $N_m=7$) and
minimal GMSB (mGMSB) with one pair of
${\bf 5} \oplus \overline{\bf 5}$ (corresponding to $N_m=1$).
For each GMSB models, $F_S/M$ has been suitably chosen
to give the same gluino mass.
We can see a sharp contrast with mGMSB mass spectrum
in which some of sfermions are heavier than corresponding
gauginos, in particular, neutralino is the NLSP.
The sparticle mass spectra of type II and III are similar,
but there are sizable mass differences between the same
sparticles, $\sim 50-70$ GeV, which will be large enough
for the precision goal of the sparticle mass measurements
at future colliders such as LHC
and the International Linear Collider (ILC).

The condition for the perturbative gauge coupling unification
leads to the lower bound on the messenger scale,
$ M \gtrsim 6.2 \times 10^{11}$ GeV.
With this $M$ and $N_m=15$, all the particle mass spectra
are determined by fixing $F_S$, so that the current experimental
lower bound on sparticle masses provide us the lower bound on
the SUSY breaking scale $F_S$.
As we will discuss more detail later, the stau NLSP
is long-lived, at least, in the collider time-scale.
The current lower mass bound for stable and long-lived massive
charged particles was obtained by LEP2 experiments
\cite{LEPlimit} as $\simeq 102$ GeV.
When we apply this bound to the NLSP stau mass,
we find $F_S/M \gtrsim 17$ TeV for
$M \simeq 6.2 \times 10^{11}$ GeV and $\tan \beta=45$ for example.
It turns out that the constraint from LFV is more severe
(see Fig.~\ref{fig3}) and $F_S/M$ should be higher as $\tan \beta$ is raised.

\subsection{gravitino dark matter}
In our $\nu$-GMSB model, the gravitino is the LSP
as in all GMSB models.
Since couplings of the gravitino to particles and sparticles
are suppressed by the Planck mass, the gravitino cannot be in the
thermal equilibrium in the early universe.
In the case with stau NLSP, gravitinos are produced
through scattering and decay processes of the MSSM particles
in the thermal plasma, and the relic density of
the gravitino LSP is evaluated as \cite{TP}
\begin{eqnarray}\label{omega}
 \Omega h^2\sim0.2\left(\frac{T_R}{10^{10}{\rm
  GeV}}\right)\left(\frac{100{\rm
  GeV}}{m_{3/2}}\right)\left(\frac{M_3}{1{\rm TeV}}\right)^{2},
\end{eqnarray}
where $T_R$ is the reheating temperature after the inflation, and
$M_3$ is the running gluino mass. By appropriately fixing the
gravitino mass, the reheating temperature and sparticle spectrum, the
relic density suitable for the dark matter can be obtained.
For our numerical examples in Table~\ref{table1} and \ref{table2},
$T_R\sim 10^5$ GeV to obtain the current dark matter
relic abundance $\Omega h^2 \simeq 0.11$ \cite{WMAP}.

\subsection{stau NLSP phenomenology}
In the model the stau is NLSP and it decays to tau and gravitino
LSP. This stau decay is of particular interests for the collider
phenomenology. The lifetime of stau NLSP is estimated as
\begin{eqnarray}
 \tau_{\tilde{\tau}} \sim 10^{-2} {\rm sec}
 \times\left(\frac{100{\rm GeV}}{m_{\tilde{\tau}_1}}\right)^5
 \left(\frac{m_{3/2}}{1{\rm MeV}}\right)^2.
\end{eqnarray}
For the values given in the Table~\ref{table1} and \ref{table2},
the lifetime of the stau is found to be
\begin{eqnarray}
\tau_{\tilde{\tau}}\sim 2.3\times 10^{-3}~{\rm sec},~~
\tau_{\tilde{\tau}}\sim 8.5\times 10^{-3}~{\rm sec}
\end{eqnarray}
for $\tan\beta=10$ and $\tan\beta=45$, respectively.
The stau lifetime is short enough not to cause
any cosmological problems, in particular,
for big bang nucleosynthesis.

On the other hand, for the stau lifetime around $10^{-3}$ sec, the
decay length well exceeds the detector size of the LHC and the ILC,
and the NLSP decay takes place outside the detector. In this case,
there have been interesting proposals \cite{sWIMPcollider} for
ways to trap long-lived NLSPs outside the detector, when the NLSP
is a charged particle (like the stau in our model). Detailed studies of
the
NLSP decay may provide precise measurements of the gravitino mass
and the four dimensional Planck mass. Very recently, the possibility of
observing long-lived NLSPs inside the detector has been investigated
\cite{Ishiwata:2008tp} for lifetimes of the NLSP $\tau
\lesssim 10^{-3}$ sec. The stau NLSP in our model is a good example
of this situation.

\section{GMSB with two {\bf 24}-dim. messengers}
In this section, we discuss implications of $\nu$-GMSB with two {\bf
24}-dimensional fields for type III seesaw as messengers. Recall that
the minimum number of {\bf 24} fields for fitting neutrino oscillation
data is two. The details of the discussion for this case is very similar
to the three
{\bf 24} case. The Dynkin index for two {\bf 24}-dimensional messenger
fields is $N_m=10$. Due to this large Dynkin index, at the messenger
scale, there is a hierarchy between the gaugino mass and sfermion mass
as before although the hierarchy in this case is less.
 For example, the ratio of the right-handed
slepton mass and Bino is found to be $m_{\tilde{e^c}}^2(M)/M_1^2(M)
= 1.2/N_m=0.12$ at the messenger scale $M$. Therefore, stau is still
most likely to be the NLSP as in the case of three {\bf 24} type III
seesaw.


To find the lower bound on the messenger scale from the gauge
coupling unification, we consider the evolution of the $\alpha_1$ in
the presence of two {\bf 24} fields, and find \bea
 M \equiv \lambda \langle S \rangle
 \gtrsim 3.4 \times 10^9 \; {\rm GeV},
\eea
in order to keep the perturbative gauge coupling unification,
say, $\alpha^{-1} (M_U) \gtrsim 1$.
This implies that the gravitino mass in this case is
\begin{eqnarray}
  m_{3/2} \gtrsim 1-10 \; {\rm keV}.
\end{eqnarray}
Since the gravitino mass can be very low in this case, i.e. $m_{3/2}
={\cal O}$(10 keV), it is a suitable candidate for the
warm dark matter of the Universe \cite{Warm}. According to the
formula Eq.~(\ref{omega}) for the gravitino production from thermal
plasma, this requires  the reheat temperature $T_R \sim 1$ TeV
 to provide the correct relic density for the gravitino dark matter.

The type III seesaw mechanism with two {\bf 24}-dim. messengers
predicts one massless light neutrino eigenstate. Since our numerical
fitting for neutrinos given in Eq.~(\ref{numass}) also assumes one
massless neutrino, the same analysis is applicable to the present case,
except we use $N_m=10$ and also different inputs for $F_S/M$.

Using the $Y_\nu$ textures for both normal- and
inverted-hierarchical cases (normal hierarchy case given in Eq.
(16)), we plot in Fig.~\ref{fig3} and \ref{fig4}, the branching
ratio $BR (\mu \to e \gamma)$ as a function of the $\Sigma$ mass $M$
and compare it with the current experimental bound, $BR (\mu \to e
\gamma) \leq 1.2 \times 10^{-11}$. In the same way, we obtain the
branching ratio for LFV tau decay $\tau\to e+\gamma$. Since the
$Y_\nu$ texture we have used here is the same as in the three {\bf
24} model, we arrive at the same result of Eq.~(\ref{ratio}) for the
ratio $BR (\tau \to \mu \gamma)/BR (\mu \to e \gamma)$.
%
%

As far as the sparticle spectrum goes, in Table~\ref{table3}, we
present examples of sparticle and Higgs mass spectra for $\tan
\beta=$10 and 45. Here we have chosen suitable inputs for $F_S/M$ to
be consistent with the current sparticle and Higgs mass bounds. In
$\tan \beta=10$ case, the resultant lightest Higgs boson mass is at
the current lower bound $m_h=114$ GeV \cite{LEP2}, while the stau
NLSP mass is at the current lower bound $m_{\tilde{\tau}_1} \simeq
102$ GeV \cite{LEPlimit} in $\tan \beta=45$ case. As Higgs mass
increases, we need to increase the $F_S/M$ value although one needs
to do more fine-tuning to get the right $Z$-mass.


Turning now to the stau NLSP, for the parameter values given in the
Table~\ref{table3}, the lifetime of the
stau is found to be
\begin{eqnarray}
\tau_{\tilde{\tau}}\sim 1.8 \times 10^{-7}~{\rm sec},~~
\tau_{\tilde{\tau}}\sim 4 \times 10^{-6}~{\rm sec}
\end{eqnarray}
for $\tan\beta=10$ and $\tan\beta=45$, respectively. Note that the
stau life is much shorter than the three {\bf 24} case due to the
fact that the SUSY breaking parameter $F_S$ can be much lower for
the two {\bf 24} case. It therefore presents a more favorable
possibility for detection
 inside the LHC detector \cite{Ishiwata:2008tp}.

\section{Conclusions}
In conclusion, we have studied the implications of the type III
seesaw mechanism for neutrino masses in an extension of SUSY SU(5)
for supersymmetry breaking phenomenology. We use the extra {\bf 24}
dim. fields needed for implementing type III seesaw as messengers
that transmit SUSY breaking from the hidden to visible sector via
the gauge forces. We find that this kind of GMSB models, specially with
three type III seesaw {\bf 24} fields, are unique
in the sense that they considerably narrow the messenger scale and
hence the SUSY breaking parameter $F_S$ compared to other GMSB
models. This allows us to make two important predictions: the cold
dark matter is a stable gravitino LSP with mass in the range of 1-10
MeV and a light stau as the NLSP with mass in the range of 200 GeV.
The stau NLSP can decay to gravitino with lifetime longer than
typical collider time scales and can be detectable at LHC. We also
find that the characteristic sparticle spectrum of this model is
different from other GMSB models which in principle can be used to
test the model. We repeat the same discussion for the case of two
{\bf 24} messengers case. The limits on the messenger mass is
clearly less restrictive in this case. As a result, the gravitino
mass can be lower and hence it can be a warm dark matter. The stau
NLSP life time is in the much more favorable range for detection at
LHC than the three {\bf 24} case.

\acknowledgments
The work of R.N.M. is supported by the National Science Foundation
Grant No. PHY-0652363. The work of N.O. is supported in part by the
Grant-in-Aid for Scientific Research from the Ministry of Education,
 Science and Culture of Japan (No. 18740170).
The work of H.B.Y. is supported by the National Science Foundation
 under Grant No. PHY-0709742. H.B.Y acknowledges the Maryland Center
 for Fundamental Physics for its hospitality during the completion
 of this work.


\newpage

\begin{table}[t]
\begin{ruledtabular}
\begin{tabular}{c|ccc}
Mediation      & Type III     & Type II       & mGMSB \\
\hline
$M  $        &  $10^{12}$     & $10^{12}$  & $10^{12}$ \\
$F_S/M$        &  15 TeV    &  31.93 TeV   &  214.2 TeV\\
\hline
$m_h$                    & 116       & 116  & 117  \\
$m_A$                    & 877       & 933  & 1394\\
$m_H$                    & 877       & 933  & 1395 \\
$m_{H^{\pm}}$            & 880       & 937  & 1397\\
\hline
$m_{\tilde{\chi}^{\pm}_{1,2}}$ & 568, 784  & 567, 813  &  557, 1090\\
$m_{\tilde{\chi}^0}$     & 299, 563, 769, 786  & 298, 564, 801, 816     & 288, 567, 1085, 1091 \\
$m_{\tilde{g}}$   & $\underline{1578} $    & $\underline{1578} $    & $\underline{1578} $   \\
\hline
$m_{{\tilde{u},\tilde{c}}_{1,2}}$  & 1357, 1394    & 1404, 1452   & 1832, 1960\\
$m_{\tilde{t}_{1,2}}$    & 1132,1342  &    1163, 1387   &  1446, 1818\\
\hline
$m_{{\tilde{d},\tilde{s}}_{1,2}}$    &  1343, 1402  & 1389, 1460    & 1800, 1969 \\
$m_{\tilde{b}_{1,2}}$    & 1307, 1336   & 1356, 1381   & 1782, 1808\\
\hline
$m_{\tilde{\nu}_{1,2,3}}$ &429,429,429 &487,487,486&883,883,881\\
\hline
$m_{{\tilde{e},\tilde{\mu}}_{1,2}}$   & 228, 439     & 273, 496     & 551, 890  \\
$m_{\tilde{\tau}_{1,2}}$  & 222, 440  & 267, 496 & 543, 888 \\
\hline
$m_{3/2}$&3.55 MeV&7.57 MeV&50.8 MeV\\
\hline
NLSP&stau&stau&neutralino
\end{tabular}
\end{ruledtabular}
\caption{
Sparticle and Higgs boson mass spectra (in units of GeV)
in the type III, type II and the mGMSB,
for $\tan \beta =10$.
} \label{table1}
\end{table}

\begin{table}[t]
\begin{ruledtabular}
\begin{tabular}{c|ccc}
Mediation      & Type III     & Type II       & mGMSB \\
\hline
$M $        &  $10^{12}$     & $10^{12}$  & $10^{12}$ \\
$F_S/M$        &  24 TeV    &  51.1 TeV    &  342.6 TeV\\
\hline
$m_h$                    & 119       & 119  & 120  \\
$m_A$                    & 941       & 1009  & 1556\\
$m_H$                    & 941       & 1009  & 1556 \\
$m_{H^{\pm}}$            & 945       & 1012  & 1558\\
\hline
$m_{\tilde{\chi}^{\pm}_{1,2}}$ & 914, 1134 & 913, 1181 &  892, 1607\\
$m_{\tilde{\chi}^0}$     & 484, 908, 1126, 1139     & 482, 908, 1175, 1186     & 465, 891, 1610, 1613 \\
$m_{\tilde{g}}$   & $\underline{2419} $    & $\underline{2419} $    & $\underline{2419} $   \\
\hline
$m_{{\tilde{u},\tilde{c}}_{1,2}}$  & 2064, 2128    & 2143, 2223   & 2843, 3051  \\
$m_{\tilde{t}_{1,2}}$              & 1748,1965      &    1799, 2040      &  2262, 2720\\
\hline
$m_{{\tilde{d},\tilde{s}}_{1,2}}$    &  2043, 2138  & 2119, 2233    & 2791, 3062 \\
$m_{\tilde{b}_{1,2}}$    & 1888, 1956   & 1954, 2031   & 2543, 2714\\
\hline
$m_{\tilde{\nu}_{1,2,3}}$ &679,679,656        &770,770,741   &1397,1397,1328\\
\hline
$m_{{\tilde{e},\tilde{\mu}}_{1,2}}$& 358, 686& 430, 777& 875, 1403  \\
$m_{\tilde{\tau}_{1,2}}$   & 207, 677  & 269, 757 & 609, 1336 \\
\hline
$m_{3/2}$&5.69 MeV&12.1 MeV&81.2 MeV\\
\hline
NLSP&stau&stau&neutralino
\end{tabular}
\end{ruledtabular}
\caption{ Sparticle and Higgs boson mass spectra (in units of GeV)
 in the type III, type II and the mGMSB, for $\tan \beta =45$.
} \label{table2}
\end{table}
\begin{table}[t]
\begin{ruledtabular}
\begin{tabular}{c|cc}
$\tan \beta$  & $10$    & $45$  \\
\hline
$M  $         &  $3.4 \times 10^9$   & $3.4 \times 10^9$  \\
$F_S/M$       &  $18.4 $ TeV           & $25.93$ TeV         \\
\hline
$m_h$           & 114       & 116   \\
$m_A$           & 669       & 646   \\
$m_H$           & 669       & 646   \\
$m_{H^{\pm}}$   & 674       & 652   \\
\hline
$m_{\tilde{\chi}^{\pm}_{1,2}}$ &  455, 613  & 649, 793   \\
$m_{\tilde{\chi}^0}$     & 244, 453, 588, 614  & 348, 647, 772, 795 \\
$m_{\tilde{g}}$   &  1314  & 1794  \\
\hline
$m_{{\tilde{u},\tilde{c}}_{1,2}}$  &  1111, 1137   & 1507, 1546 \\
$m_{\tilde{t}_{1,2}}$    &  946, 1113   &  1298, 1458 \\
\hline
$m_{{\tilde{d},\tilde{s}}_{1,2}}$  & 1101, 1145   & 1493, 1555 \\
$m_{\tilde{b}_{1,2}}$    & 1075, 1097   & 1386, 1446 \\
\hline
$m_{\tilde{\nu}_{1,2,3}}$ &  327, 327, 327  & 459, 459, 447 \\
\hline
$m_{{\tilde{e},\tilde{\mu}}_{1,2}}$   & 170, 339 & 235, 469  \\
$m_{\tilde{\tau}_{1,2}}$  & 164, 340  & 102, 477  \\
\hline
$m_{3/2}$ & 14.8 keV  & 20.9 keV\\
\hline
NLSP&stau&stau
\end{tabular}
\end{ruledtabular}
\caption{ Sparticle and Higgs boson mass spectra (in units of GeV)
of the model with two $\bf 24$ messengers for $\tan \beta =10$ and
45. }\label{table3}
\end{table}
\newpage

\begin{figure}[t]
\includegraphics[scale=1.2]{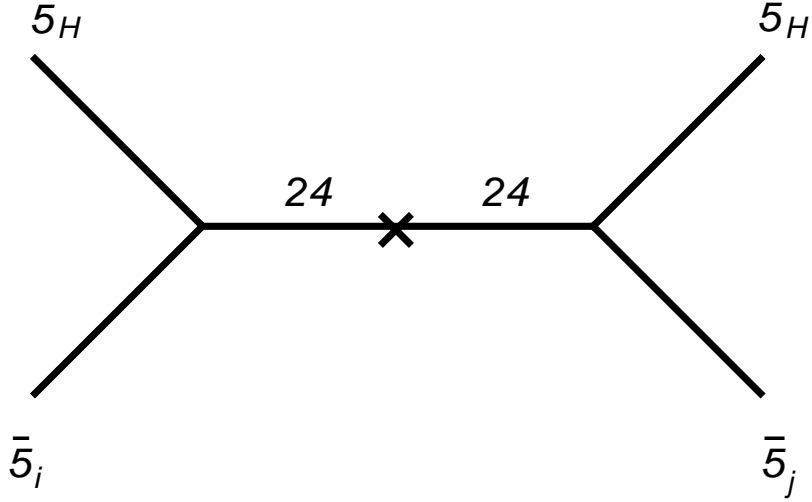}
\caption{
The diagram of type III seesaw mechanism.
}\label{fig1}
\end{figure}
\begin{figure}[t]
\includegraphics[scale=1.2]{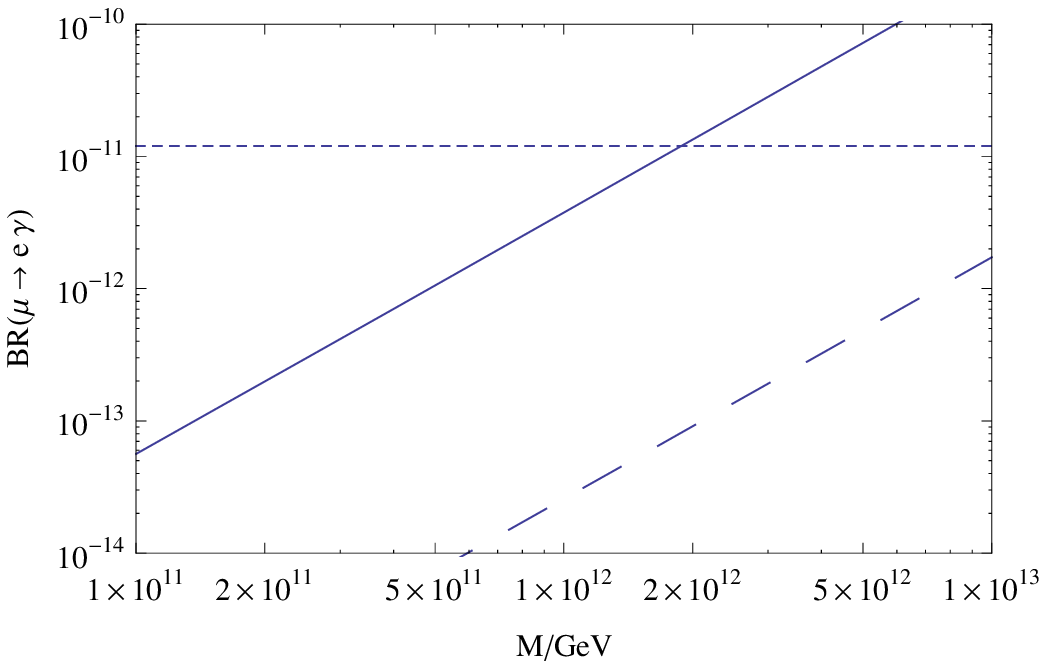}
\caption{ The branching ratio of the muon LFV decay as a function of
$M$, for $\tan \beta = 10$ of three $\bf 24$ model. Here we have
fixed $F_S/M=15$ TeV (see Table~I). The solid and dashed lines
correspond to the normal- and inverted-hierarchical cases,
respectively, while the dotted horizontal line is the current
experimental upper bound $BR(\mu \to e \gamma) \leq 1.2 \times
10^{-11}$. }\label{fig2}
\end{figure}
\begin{figure}[t]
\includegraphics[scale=1.2]{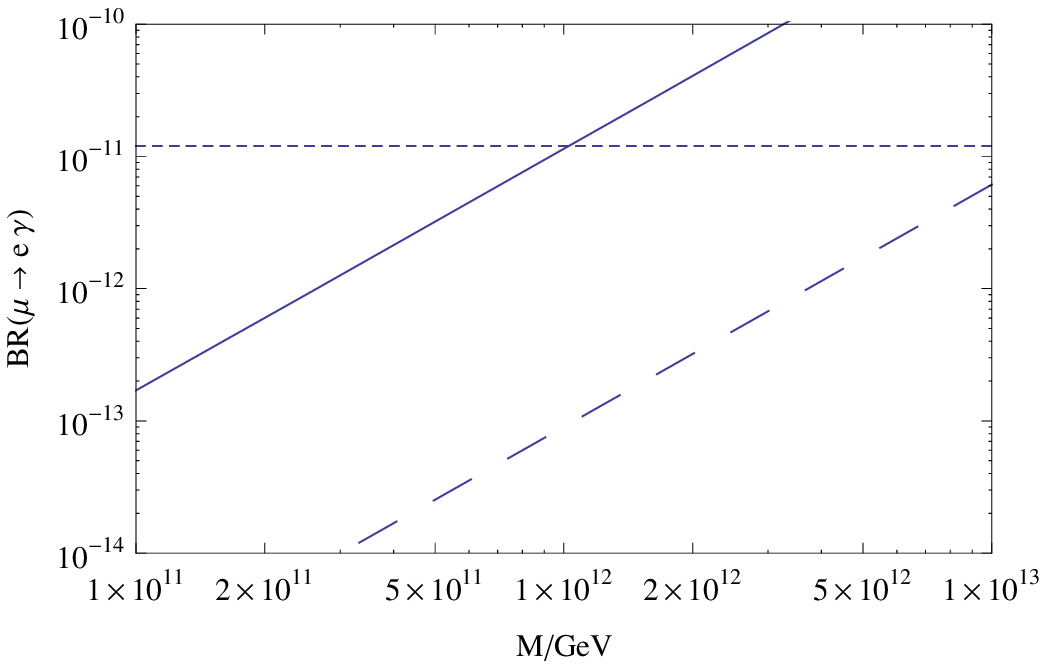}
\caption{
Same as Fig.~\ref{fig2}, but for $\tan \beta = 45$.
Here we have fixed $F_S/M=24$ TeV (see Table~II).
}\label{fig3}
\end{figure}
\begin{figure}[t]
\includegraphics[scale=1.2]{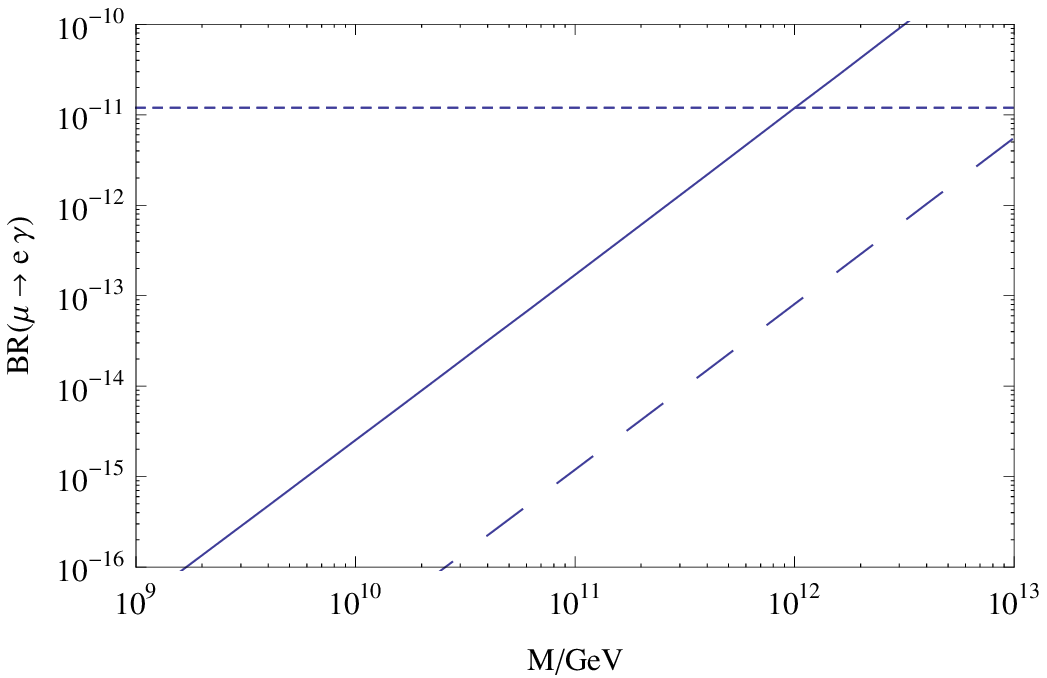}
\caption{ The branching ratio of the muon LFV decay as a function of
$M$, for $\tan \beta = 10$ of two-$\bf 24$ model. Here we have fixed
$F_S/M=18.4$ TeV (see Table \ref{table3}). The solid and dashed
lines correspond to the normal- and inverted-hierarchical cases,
respectively, while the dotted horizontal line is the current
experimental upper bound $BR(\mu \to e \gamma) \leq 1.2 \times
10^{-11}$. }\label{fig4}
\end{figure}
\begin{figure}[t]
\includegraphics[scale=1.2]{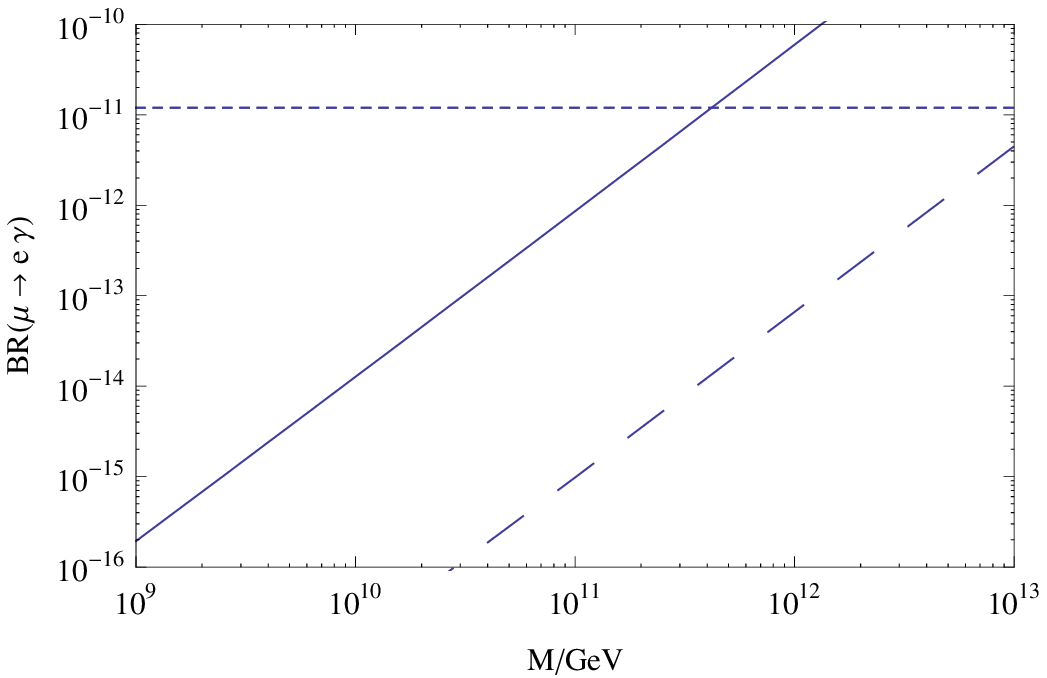}
\caption{Same as Fig.~\ref{fig4}, but for $\tan \beta = 45$.
Here we have fixed $F_S/M=25.93$ TeV (see Table \ref{table3}).
}\label{fig5}
\end{figure}

\end{document}